\documentclass[aps,twocolumn,prd,showpacs,nofootinbib,showkeys,superscriptaddress,preprintnumbers]{revtex4}

\usepackage{graphicx,floatflt}
\usepackage{amsmath}
\usepackage{epsfig}

\newcommand{\dd}{\mathrm{d}} 
\newcommand{\Mpl}{M_\mathrm{pl}} 
\newcommand{\mpl}{m_\mathrm{pl}} 

\newcommand{\epsone}{\epsilon_1}

\newcommand{\rr}{\mathrm}

\begin{document}

\preprint{ULB-TH/10-22, CP3-10-24}

\title{Hybrid inflation along waterfall trajectories}

\author{S\'ebastien Clesse}
\email{seclesse@ulb.ac.be}
\affiliation{Service de Physique Th\'eorique, Universit\'e Libre de Bruxelles, 
CP225, Boulevard du Triomphe, 1050 Brussels, Belgium}
\affiliation{Institute of Mathematics and Physics, Center for Cosmology, Phenomenology and Particle Physics, Louvain University, 2 chemin du cyclotron, 1348 Louvain-la-Neuve, Belgium}

\date{\today}

\begin{abstract}
We identify a new inflationary regime for which more than 60 e-folds are generated classically during the waterfall phase occuring after the usual hybrid inflation.  By performing a bayesian Monte-Carlo-Markov-Chain analysis, this scenario is shown to take place in a large part of the parameter space of the model.   When this occurs, the observable perturbation modes leave the Hubble radius during waterfall inflation.  The power spectrum of adiabatic perturbations is red, possibly in agreement with CMB constraints.  A particular attention has been given to study only the regions for which quantum backreactions do not affect the classical dynamics.  Implications concerning the preheating and the absence of topological defects in our universe are discussed.    

\end{abstract}
\pacs{98.80.Cq}
\maketitle

\section{Introduction}

In the last two decades, precise measurements of Cosmic Microwave Background (CMB) anisotropies have provided strong arguments in favor of a phase of accelerated expansion in the early universe.  These arguments have lead to propose a large number of inflationary models, most of them based on one or more scalar fields slowly rolling along an associated potential, and whose predictions on the power spectrum of scalar initial perturbations agree with best the measurements of CMB anisotropies.   Inside the jungle of models, the hybrid class is particularly promising because easily embedded in a high energy framework like supersymmetry, supergravity, or grand unified theories \cite{Mazumdar:2010sa}.   Contrary to power law chaotic models \cite{Martin:2010kz}, the energy scale of hybrid inflation can be low, and the model do not need super-plankian values of the fields.  Moreover, it is shown in~\cite{Clesse:2009ur,Clesse:2008pf,Clesse:2009zd} that initial field values in hybrid inflation do not suffer a fine-tuning problem, unlike previously thought~\cite{Tetradis:1997kp,Mendes:2000sq}.   Nevertheless, the original hybrid model is usually considered as a toy model, because its scalar power spectrum calculated in the 1-field slow-roll approximation exhibits a slight blue tilt, which is disfavored by CMB experiments~\cite{Martin:2006rs}.   In this paper, we discuss the validity of that calculation.  We show that in a large part of the parameter space, the last $60$ e-folds of inflation, relevant for the calculation of the observable scalar power spectrum, are actually realized in a non-trivial way during the waterfall phase, after crossing the instability point.  

The original hybrid model of inflation was proposed in
Refs.~\cite{Linde:1993cn,Copeland:1994vg}. Its potential reads
\begin{equation} \label{eq:potenhyb2dNEW}
V(\phi,\psi) = \Lambda^4 \left[  \left( 1 - \frac{\psi^2}{M^2} \right)^2 
+ \frac{\phi^2}{\mu^2} + \frac{\phi^2 \psi^2}{\nu^4}\right] .
\end{equation}
The field $\phi$ is the inflaton, $\psi$ is the auxiliary
Higgs-type field, and
$M,\mu,\nu$ are three mass parameters. 
Inflation is assumed to be realized in the
false-vacuum along the valley $\langle\psi\rangle=0$.  In the usual description, inflation ends due to a tachyonic instability soon after the inflaton reaches a critical value $ \phi_{\rr c} =   \sqrt 2 \nu^2 / M $.  From this point, the classical system would evolve quickly toward one of 
its true minima $\langle\phi\rangle=0$, and $\langle\psi\rangle=\pm
M$ whereas in a realistic scenario one expects the tachyonic
instability to trigger a preheating era~\cite{Kofman:1997yn,
  Garcia-Bellido:1997wm, Felder:2000hj, Felder:2001kt, Copeland:2002ku, Senoguz:2004vu, Micha:2004bv,
  Allahverdi:2007zz}.

To determine the scalar power spectrum at the end of inflation, it is common usage to restrict the dynamics to the effective one-field potential along the valley $\psi = 0$
\begin{equation} \label{eq:effpot}
  V_{\rr{eff}} (\phi) = \Lambda^4 \left[ 1 + \left( \frac \phi \mu \right)^2  \right] ,
\end{equation}   
and to assume that inflation ends abruptly once instability point is reached.   Under these hypotheses, the primordial scalar power spectrum can be easily derived in the slow-roll approximiation \cite{Leach:2002ar}.
It is nearly scale invariant, with a spectral tilt
\begin{equation} \label{eq:nsslowroll}
n_{\mathrm s} -1 = - 2 \epsilon_{\rr 1*} - \epsilon_{\rr 2*},
\end{equation}
where a star means that the quantity is evaluated when the pivot mode $k_{\rr 0}$ leaves the Hubble radius, that is when $k_{\rr 0} = a H$. 

However, let us remind that this formalism is based on the belief that inflation ends immediately when the instability point $\phi_{\rr c}$ is reached.  For original hybrid model and its effective potential (Eq.~\ref{eq:effpot}), the slow-roll parameters read~\cite{Martin:2006rs}\footnote{Throughout the paper, $\mpl$ denotes the physical Planck
  mass, and $\Mpl$ stands for the reduced Planck mass $\Mpl \simeq 0.2
  \mpl\simeq 2.4\times 10^{18}$~GeV.}
\begin{equation} \begin{split}
\epsilon_{\rr 1} & = \frac{\mpl^2}{16 \pi} \left( \frac{V'}{V}  \right)^2\\
& = \frac {1} {4 \pi} \left( \frac{\mpl}{\mu}  \right)^2 \dfrac{\left(\dfrac{\phi }{ \mu} \right)^2}{\left[1 + \left( \dfrac{\phi}{ \mu} \right)^2  \right]^2} \ ,  \\
\epsilon_{\rr 2} & = \frac{\mpl^2}{4 \pi} \left[\left( \frac{V'}{V}  \right)^2 - \frac{V''}{V}  \right]\\
& =  \frac {1} {2 \pi} \left( \frac{\mpl}{\mu}  \right)^2 \dfrac{\left(\dfrac{\phi }{ \mu }\right)^2-1}{\left[1 + \left( \dfrac{ \phi}{ \mu} \right)^2  \right]^2} \ ,
\end{split} \label{eq:slowrollparams}
\end{equation}
where a prime denotes the derivative with respect to the inflaton field. Throughout the paper, we restrict our analysis to the case in which inflation occurs in the small field phase ($\phi \ll \mu$).  In this regime $ \epsilon_{\rr 1}$ is extremely small, $\epsilon_{\rr 2}$ is negative and is the dominant contribution to the spectral tilt.  Thus the scalar spectral index is generically larger than one and the spectrum is blue, which is disfavored by WMAP7 observations~\cite{Larson:2010gs}.  

In this paper, we relax the hypothesis of instantaneous end of inflation at the instability point  
and look for trajectories performing a phase of inflation during the waterfall \cite{Mazumdar:2003jv}.  Tachyonic preheating is not triggered during this phase because the effective mass of the adiabatic field is small compared to the expansion term.  
More precisely,  the exponential growth of perturbation modes, characteristic of tachyonic preheating, is avoided because the Hubble expansion term is dominating the equation governing the linear perturbations of $\psi$.
  In that case, both analytical investigations and lattice simulations describing the tachyonic preheating do not apply.  

In this paper, the classical dynamics is investigated both numerically and analytically by using the adiabatic field formalism \cite{Gordon:2000hv}.  The whole potential parameter space is explored using a Monte-Carlo-Markov-Chains (MCMC) method.  Regions for which much more than 60 e-folds are realized after instability are shown to be generic.    In such cases, observable modes leave the Hubble radius during waterfall inflation and a modification of the predicted scalar spectral index is expected.  For adiabatic perturbations, the power spectrum is actually generically red.  

The potential is very flat near the critical instability point and quantum backreactions could dominate the classical dynamics.  Therefore, a particular attention has been given to consider only trajectories not affected by quantum stochastic effects.   These comprise both the quantum backreactions of the adiabatic and the entropic transverse field.  The classical regime is valid only if classical jumps of the adiabatic field are larger than its quantum fluctuations and if quantum diffusion of the transverse field \cite{GarciaBellido:1996qt}  do not increase too much the spread of its probability distribution.



For a short phase of waterfall inflation, some problems are put in evidence in \cite{GarciaBellido:1996qt}.   Inflating topological defects can induce large-scale perturbations and primordial black holes can be formed after inflation.  When the waterfall phase is much longer, we argue that these problems are naturally avoided.  Indeed, topological defects are so strongly diluted during waterfall inflation that they do not affect the observable universe.  Primordial black holes are expected to form when fractional density perturbations occurring at the phase transition reenter the horizon.  Thus they affect the observable universe only if waterfall inflation lasts less than typically $60$ e-folds.  This is not the case here.

The paper is organized as follows:  section 2 is dedicated to the dynamics inside the valley, before the instability point is reached.   It is shown that classical oscillations of the waterfall field are quickly dominated by its quantum fluctuations.  In section 3, we show that much more than $60$ e-folds can be realized along a waterfall trajectory, i.e. after crossing the critical instability point.   In section 4 the generic character of this effect is studied.  The dependences on the potential parameters and initial conditions are determined by using a MCMC method.   In the conclusion, important implications  for hybrid models (e.g. on the formation of topological defects) are discussed.

\section{Field dynamics before instability}

Given an arbitrary set of initial conditions, two classical behaviors are possible.  Either the trajectory falls through one of the global minima of the potential without inflating.  Either it reaches the nearly flat valley along $\psi = 0 $ and slow-roll inflation can occur.   If the valley is reached,  trajectories are characterized by damped oscillations in the transverse direction (orthogonal to the valley).  After some oscillations, slow-roll regime begins along the bottom of the valley and a large number of e-folds is realised.   

At the critical point of instability $\phi_{\rr c}$, only a small transverse displacement allows inflation to end with a waterfall phase.  The two competing processes able to cause this displacement from the $\psi = 0$ valley line are the remaining classical transverse oscillations and the quantum fluctuations of the auxiliary field.   In this section, it is shown that oscillations own generically an amplitude so small that they are dominated by quantum fluctuations of the auxiliary field.

Quantum fluctuations are typically of the order $\Delta \psi_{\rr{qu}} \simeq H /2 \pi$.
Primordial nucleosynthesis fixes a reasonable lower bound on the energy scale of inflation, and thus on $H$ through the Friedmann-Lemaitre equations (Eq.~\ref{eq:FLtc12field}).   On the other hand, measurements of the primordial scalar power spectrum amplitude,
 \begin{equation}
\mathcal P(k=0.002/\rr{Mpc})\simeq 2.43 \times10^{-9} = \frac{H_* ^2} {\pi \mpl^2 \epsilon_{\rr 1 *}}, 
\end{equation}
with $ \epsilon_{\rr 1} \lesssim 0.1 $,  allow to fix a higher bound on $H$.  
One can thus determine the range of transverse quantum oscillations, $10^{-30} \mpl \lesssim \Delta \psi_{\rr{qu}} \lesssim 10^{-6} \mpl $.


To study the classical oscillations, the two-field dynamics needs to be integrated.  For a flat Friedmann-Lema\^{\i}tre-Robertson-Walker metric, the
equations governing the two-field dynamics are the
Friedmann-Lema\^{\i}tre equations,
\begin{equation} \label{eq:FLtc12field}
\begin{split}
H^2 &= \frac {8\pi }{3 \mpl^2}  \left[ \frac 1 2 \left(\dot
\phi^2 + \dot \psi^2 \right)  + V(\phi,\psi) \right], \\
\frac{\ddot a }{a} &= \frac {8\pi}{3 \mpl^2} \left[ - \dot \phi^2
- \dot \psi^2 + V(\phi,\psi ) \right]~,
\end{split}
\end{equation}
as well as the Klein-Gordon equations 
\begin{equation} \label{eq:KGtc2field}
\begin{split}
&\ddot \phi + 3 H \dot \phi + \frac {\partial
V(\phi,\psi)}{\partial \phi} = 0~, \\
&\ddot \psi + 3 H \dot \psi + \frac {\partial 
V(\phi,\psi)}{\partial \psi} = 0~.
\end{split}
\end{equation}
where $H \equiv \dot a/a$ is the
Hubble parameter, $a$ is the scale factor and a dot denotes derivative
with respect to cosmic time.   

In the regime of small classical oscillations $\psi \ll M$, at inflaton values sufficiently larger than the critical one but still in the small field phase, that is $\phi_{\rr c} \ll \phi \ll \mu$, the potential is well approximated by
\begin{equation}
V(\phi,\psi) \simeq \Lambda^4 \left[ 1 + \frac{\phi^2}{\mu^2} + \frac{\phi^2 \psi^2 }{\nu^4}  \right] \ .
\end{equation}

We can assume that the inflaton field is slow-rolling along the valley $\psi = 0$, such that  
 \begin{equation} \label{eq:Hevol}
H \simeq \frac{1}{\sqrt 3 \Mpl} \sqrt{\Lambda^4 (1+ \phi^2 / \mu^2)} \ .
\end{equation}   
In a short time scale, the inflaton $\phi$, and thus the Hubble parameter $H$, can be assumed to be constant.
As a consequence, the Klein-Gordon equation for the auxiliary field now read
\begin{equation}
\ddot \psi + 3  \dot \psi \frac{1}{\sqrt 3 \Mpl} \sqrt{\Lambda^4 (1+ \phi^2 / \mu^2)} + \frac{2 \Lambda^4 \phi^2} {\nu^4} \psi = 0 \ .
\end{equation}
It has a simple oscillating solution with exponentially decreasing amplitude
\begin{equation}
\begin{aligned}
\psi(t)  & =  \ \rr e ^{-  \frac{3}{2 \sqrt 3 \Mpl} \sqrt{\Lambda^4 (1+ \phi^2 / \mu^2)} t}  \\ 
 & \left[ C_{\rr 1} \  \rr e^{-\frac 3 2 \frac{1}{\sqrt 3 \Mpl} \sqrt{\Lambda^4 (1+ \phi^2 / \mu^2)} t \sqrt{1-8 \Mpl^2 \phi^2 / (3 \nu^4)} } \right.    \\
  & \left.   + C_{\rr 2} \  \rr e ^{ \frac{3}{2 \sqrt 3 \Mpl} \sqrt{\Lambda^4 (1+ \phi^2 / \mu^2)} t \sqrt{1- 8 \Mpl^2 \phi^2 / (3 \nu^4) } }  \right] , 
\end{aligned}
\end{equation}
where $C_{\rr 1} $ and $C_{\rr 2}$ are two integrating constant fixed by initial conditions.  As an example, it takes about $N \simeq 40 $ e-folds
for initial oscillations of amplitude $A\simeq 10^{-3} \mpl $ to be reduced by a factor of $10^{27}$ at the minimal level of quantum fluctuations.   

Given this time scale, the assumption that $\phi$ is constant can be justified a posteriori.  Indeed, in the slow-roll approximation, straightforward manipulations ~\cite{Martin:2006rs} give 

\begin{equation} \label{eq:Ndephi}
N(\phi) = \frac{\mu^2}{4 \Mpl^2} \left[ \left(\frac{\phi_{\rr i}}{\mu} \right)^2 - \left( \frac{\phi}{\mu}  \right)^2 - 2 \ln \left( \frac{\phi}{\phi_{\rr i}} \right) \right],
\end{equation}
where $\phi_{\rr i}$ is the initial inflaton value.  Therefore, if $\phi \ll \mu$, classical oscillations become dominated by quantum fluctuations in a range of inflaton value 
\begin{equation}
\frac{\Delta \phi }{ \phi} \simeq - \rr e^{- \frac{2 \Mpl^2}{\mu^2} N } \  ,
\end{equation} 
which is typically very small for a nearly flat valley.   
Thus the classical oscillations of $\psi$ are expected to be dominated by quantum fluctuations after a very small range of variation for $\phi$.    

To go beyond this approximation and determine, for the full potential, how generic are trajectories whose classical oscillations become dominated by transverse quantum fluctuations,  the classical 2-field dynamics have been integrated numerically.  We have followed the method used in~\cite{Clesse:2009ur} and run an identical Monte-Carlo-Markov-Chains analysis
of the 7D space of initial field values, initial velocities and potential parameters.   The result of this analysis is that around $99.9 \% $ of trajectories trapped inside the valley perform transverse oscillations whose amplitude is below the most restrictive level of transverse quantum fluctuations $\Delta \psi_{\rr{qu}} \sim 10^{-30} \mpl$.  

From these considerations, at instability, the slight transverse displacement essential for a waterfall phase to take place is not supplied by classical oscillations of the auxiliary field but by its quantum fluctuations.  In the following sections, the waterfall phase will be studied classically taking initial values $\phi_{\rr i} = \phi_{\rr c}$ and $\psi_{\rr i} = \Delta \psi_{\rr {qu}} $.   We will assume initial field velocities given by the slow-roll approximation.   Actually, due to the slow-roll attractor, different choices of initial velocities only marginally influence the resulting waterfall dynamics.

\section{How hybrid inflation ends}

Before to study the waterfall phase, it must be verified that the classical dynamics is valid and not spoiled by quantum backreactions of both the adiabatic and the entropic fields.  

\subsection{Quantum backreactions}


The collective evolution of the fields can be described by the adiabatic field $\sigma$, defined hereafter in Eq.~(\ref{eq:adiabaticfield}).  It is very light and its classical evolution is valid if classical jumps are larger than the quantum fluctuation scale, that is
\begin{equation}
\Delta \sigma _{\rr{cl}} = \frac{\dot \sigma}{H} > \Delta \sigma_{\rr{qu}} \simeq \frac{H}{2 \pi} \ .
\end{equation}
which is equivalent to 
\begin{equation} \label{eq:hardprior}
\epsilon_{\rr 1} (\sigma) > \frac{H^2}{\pi \mpl^2} \ .
\end{equation}
Thus we pay a particular attention to only consider waterfall trajectories along which this condition remains true.  

At the critical instability point, the classical value of the transverse field is about 0, and the transverse quantum fluctuations will determine on which side the system will evolve towards. The overall dynamics remains classical due to the $\phi$ field evolution.  However, it must be ensured that the quantum backreactions of the transverse field do not push the field evolution far from the valley line $\psi = 0$.  Such effects would modify strongly the dynamics such that the waterfall phase would take place in a low number of e-folds.  In other words, it must be ensured that the spread of the probability distribution of the auxiliary field does not becomes much larger than its classical value during the waterfall.

 The coarsed-grained auxiliary field is described by a Klein-Gordon equation in which a random noise field $\xi (t)$ is added.  This term acts as a classical stochastic source term.  In the slow-roll approximation, the evolution is given by the first order Langevin equation 
\begin{equation}
\dot \psi + \frac{1}{3 H} \frac{\dd V}{\dd \psi}= \frac{H^{3/2}}{2 \pi} \xi(t) ~.
\end{equation}
which can be rewritten
\begin{equation}
\dot \psi = \frac{H^{3/2}}{2 \pi} \xi(t) + H \frac{4 \psi \Mpl^2}{M^2} \left( 1- \frac{\phi^2}{\phi_{\rr c} ^2}\right) ~.
\end{equation}
The two-points correlation function of the noise field obeys to
\begin{equation}   
\langle \xi(t) \rangle = 0,  \hspace{5mm} \langle \xi(t) \xi(t') \rangle = \delta (t-t') ~.
\end{equation}
When the dynamics is dominated by the $\phi$ evolution, this equation can be integrated exactly.  Under  a change of variable \cite{GarciaBellido:1996qt}, $x\equiv \exp \left[-2 r (N-N_{\rr{c}}) \right]$, where  
\begin{equation}
r\equiv \frac 3 2 - \sqrt{\frac 9 4 - 6 \frac{\Mpl^2}{ \mu^2}}~,
\end{equation}
and where $N_{\rr c} $ is the number of e-folds at critical point of instability, one has
\begin{equation}
\frac{\dd \psi}{\dd x} = - \frac{H^{1/2} } { 4 \pi r x} \xi(x) - \frac {4 \psi \Mpl^2 (1-x)}{2 M^2 r x}~.
\end{equation}
The exact solution to this equation is
\begin{equation}
\begin{aligned}
\psi(x) & =  C \exp \left[ C_2 x - C_2 \ln x  \right] \\ 
 & - C_1 \exp \left[ C_2 x - C_2 \ln x \right]  \\ 
 & \times \int_1^x \exp \left[ -C_2 x' + C_2 \ln x' \right] \xi(x') \dd x' ~,
\end{aligned}
\end{equation}
where $C_1 \equiv H^{1/2} /  4 \pi r  $, $C_2 \equiv 2/  M^2 r   $ and $C$ is a constant of integration.  Taking the two point correlation function and assuming an initial delta distribution for $\psi$ at $\phi \gg \phi_{\rr c}$, one can obtain 
\begin{equation} \label{eq:psi_qudist}
\langle \psi^2(x) \rangle = \frac {H^2}{8 \pi^2 r} \left[ \frac{\exp (x)}{ a x}  \right]^a  \Gamma(a,a x) \ ,
\end{equation}
where $a\equiv 4 \Mpl^2 /M^2 r$ and  $\Gamma$ is the upper incomplete gamma function.

In the following, we will consider large values of $\mu$ and relatively small values of $M$ compared to the Planck mass, such that $ r \simeq 2 \Mpl^ 2 / \mu^2$ and $a\simeq 2 \mu^2 /M^2 \gg 1$.  
At instability, $x = 1$ and one thus has
\begin{equation}  \label{eq:qufluct}
\langle \psi^2(x=1) \rangle \simeq \frac {H^2 \mu^2}{16 \pi^2 \Mpl^2} \left( \frac{\rr e M^2}{ 2 \mu^2}  \right)^{\frac{2 \mu^2}{ M^2}}  \Gamma \left(\frac{ 2 \mu^2}{M^2} , \frac{ 2 \mu^2}{M^2}  \right).
\end{equation}
By using recurrence relations as well as the asymptotic behavior of the $\Gamma$ function, one can find
\begin{equation}
\left(  \frac{\rr e}{u} \right)^u \times \Gamma(u,u) \sim \sqrt{\frac \pi 2} \frac 1 {\sqrt u} \hspace{2mm} \rr{when} \hspace{2mm} u \rightarrow \infty ~,
\end{equation} 
such that 
\begin{equation} \label{eq:variance_at_inst}
\langle \psi^2(x=1) \rangle \simeq \frac{H^2 \mu M}{32 \pi^{3/2} \Mpl^2} 
\end{equation}
For instance, for the parameter values of Fig.\ref{fig:traj_phipsi2}, $\mu = 636.4 \ \mpl, M=0.03 \ \mpl$, one obtains $ \sqrt{\langle \psi^2 \rangle} \simeq 2 H $
It will be shown later in the paper (see Sec.\ref{sec:paramspace}) that when the tachyonic preheating is triggered and forces inflation to end,  one has $x  \lesssim 1$ such that 
$\langle \psi^2(x_{\rr{end}}) \rangle \sim \langle \psi^2(x=1) \rangle $ $\sim H$.   Therefore, after the critical instability, all the time the field dynamics is governed mainly by the $\phi  $ evolution, the standard deviation of the transverse field distribution around its classical value remains at the same order.  

Notice that an identical result can be obtained using the formalism developped in \cite{Gong:2010zf,Fonseca:2010nk,Abolhasani:2010kn,Abolhasani:2010kr,Lyth:2010ch}.  Since the auxiliary field is well anchored at its minimum $\psi = 0$ before the waterfall, it can be regarded as the same as its fluctuation, $\psi = \delta \psi$.  As for the stochastic formalism, one can assume that the $\phi$ field evolves independently according to Eq.~(\ref{eq:Ndephi}).   By Fourier expanding $\delta \psi$ and neglecting non linear terms, one obtains the mode evolution equation
\begin{equation}
\delta \psi_k '' + 3 \delta \psi_k ' + \left\{  \frac{k^2}{a^2 H^2 } - 12 \frac{\Mpl^2}{M^2} \left[1 -  \rr e ^{-2 r (N-N_{\rr c})} \right] \right\} \delta \psi_k = 0 ~.  
\end{equation}
where a prime denotes derivative with respect to the number of e-folds.  Following Ref.~\cite{Gong:2010zf}, in the high frequency limit this equation can be solved in terms of the WKB approximation.  In the low frequency limit its exact solution is a combination of the Hankel functions of first and second kind.   Let introduce $k_{\rr c} \equiv a_{\rr c} H$ the mode leaving the Hubble radius at the critical instability, and $n \equiv N-N_{\rr c}$.  Then, under the assumption that $12 \Mpl^2 / M^2 \gg 1$, one finds for small scales modes near the instability $k \gg k_{\rr c} \exp (n)$, 
\begin{equation} \begin{split}
&\left|   \delta \psi_{\rr S} (k,n) \right|= \frac{H}{\sqrt{2 k}  k_{\rr c}} A \\
& \times  \exp \left(  \frac 2 3 \alpha n^{3/2} - \frac 3 2 n - \frac{1}{4} \log n \right) ~,
\end{split} \end{equation}
and for large scales modes $k \ll k_{\rr c}  \exp (n)$, 
\begin{equation} \begin{split} \label{eq:smallk}
& \left|  \delta \psi_{\rr L} (k,n) \right| =  \frac{H}{\sqrt{2 \alpha k_{\rr c} ^3}  } \\ 
 & \times \exp \left( \frac 2 3 \alpha n^{3/2} - \frac 3 2  n - \frac{1}{4} \log n \right)~,
\end{split}
\end{equation}
where $A \equiv 3^{2/3} \Gamma (2/3) \alpha^{-1/6}/(2 \sqrt \pi)$ is a typically order unity factor, and 
\begin{equation}
\alpha  \equiv \sqrt{24 r \frac{\Mpl^2 }{M^2 } }
\end{equation}
In the regime $2 r n \ll 1 $, the modes which become tachyonic satisfy 
\begin{equation}
\left(\frac k {k_{\rr c}}\right) ^2 \le \alpha^2 n \  \rr e ^{2n} ~.
\end{equation} 
In  \cite{Gong:2010zf}, it is then assumed that $\alpha \gg 1$ and $n \sim \mathcal O (1)$ to find that the quantum back-reactions from the small scales entropy perturbations dominate and force inflation to end quickly after waterfall instability.   We will be here interested by the opposite case, $\alpha \lesssim 1$.  
As shown later in Sec.\ref{sec:paramspace}, the total number of e-folds that can be realized classically between the instability and the beginning of the tachyonic preheating is larger than 60 and is roughly given by $ n \sim \mu^2 M^2 $.  Therefore the tachyonic modes stand super-Hubble during all this phase.   In that case, the variance of $\delta \psi$ is dominated by the large scale mode contribution 
\begin{equation} 
\langle \delta \psi^2 (n)  \rangle  =  \int_0 ^{k_{\rr c} \rr e^n} \frac{\dd ^3 k}{(2 \pi)^3}\delta \psi_L ^2 (k,n) ~.
\end{equation} 

One obtains just after the critical instability 
\begin{equation} \begin{split}
\langle \delta \psi^2 [n\sim \mathcal O(1)]  \rangle &\simeq  \frac{H^2}{12 \pi^2 \alpha} \exp \left( \frac 4 3 \alpha n^{3/2}  - \frac{1}{2} \log n \right) \\
& \simeq   \frac{H^2}{12 \pi^2 \alpha}
\end{split}
\end{equation}
which is identical to Eq.~(\ref{eq:variance_at_inst}) up to an order unity numerical factor\footnote{Notice that a similar result can be obtained for $n=0$, from the Eq. (2.27) and (2.30) of Ref. \cite{Gong:2010zf}.  In that case, an additional factor $\alpha^{-1/6}$ is obtained, but it can be due to the matching problem between large and small scale modes.}.

In the section \ref{sec:classical}, initial values of $\psi$ at critical instability are taken to follow a gaussian random distribution that verify Eq.~(\ref{eq:qufluct}).   From this point, the classical value of $\psi$ moves away its initial amplitude and increases such that it becomes quickly much larger than its quantum fluctuations, even if the overall dynamics is still governed mainly by the $\phi$ evolution.  Therefore, the classical dynamics is not spoiled by transverse quantum fluctuations.


\subsection{Transverse field gradient contribution}

Another effect susceptible to spoil the classical homogeneous dynamics is the backreaction due to the transverse field gradient contribution to the energy density.  

Assuming the statistical homogeneity, the mean-square value of transverse field gradient after smoothing on a length $L = 1/ a H $  is given by 
\begin{equation}
\langle |  \nabla \psi |^2 \rangle = \frac{1}{(2 \pi)^3} \int_0 ^{a H} (\dd k)^3  \left( \frac k a \right)^2 | \psi_k | ^2   ~.
\end{equation}
During the waterfall, $\psi_k$  is given by Eq.~(\ref{eq:smallk}).   After integration over the modes, one obtains
\begin{equation}
\langle |  \nabla \psi |^2 \rangle  \sim H^2  \langle | \psi |^2 \rangle  ~.
\end{equation}

Since $\langle | \psi |^2 \rangle \sim H^2$ during the waterfall in the regime of interest, the transverse field gradient contribution to the energy density is negligible compared to the potential term $ V \simeq 3 H^2 / \Mpl^2 $.   The background dynamics thus remains mostly homogeneous.


\subsection{The classical regime} \label{sec:classical}

Once the instability point is reached, trajectories deviates from the valley line and fall through one of the global minima $(\phi=0,\psi=\pm M)$.  Following \cite{Copeland:2002ku}, we can assume that the auxiliary field reacts faster than the inflaton field such that trajectories follow the ellipse defined by the minima in the $\psi$ direction
\begin{equation} \label{eq:ellipse}
\frac {\dd V(\phi,\psi)}{\dd \psi} = 0 \ \rr{ with } \  -\phi_c \le \phi \le \phi_c \Longrightarrow \frac{\psi^2}{M^2} + \frac{\phi^2}{\phi_c^2} =  1 \ .
\end{equation} 
The small-field type effective potential defined by this ellipse is nearly flat around the instability point, where its curvature is negative.  Thus a phase of inflation should be possible near the top.   The adiabatic field $\sigma$, introduced in \cite{Gordon:2000hv}, defined such that
\begin{equation} \label{eq:adiabaticfield}
\dot \sigma = \sqrt{\dot \phi^2 + \dot \psi^2  } \ ,
\end{equation}
describes the collective evolution of the fields along the classical trajectory.  One can determine its equation of motion
\begin{equation}
\ddot \sigma + 3 H \dot \sigma + V_\sigma = 0 \ ,
\end{equation}
where 
\begin{equation}
V_\sigma = \frac{\dot \phi}{\dot \sigma} \frac{\dd V}{\dd \phi}+ \frac{\dot \psi}{\dot \sigma} \frac{\dd V}{\dd \psi} \ .
\end{equation}
On the ellipse of Eq.~(\ref{eq:ellipse}), one obtains 
\begin{equation} \label{eq:potadiabatic}
V_\sigma = \Lambda^4 \frac{2 \dfrac{\phi}{\mu^2} + 4 \dfrac{\phi}{\phi_{\rr c}^2 } \left( 1 - \dfrac{\phi^2}{\phi_{\rr c}^2 } \right)  }{ \sqrt{1+ \dfrac{M^2 \phi^2}{\phi_{\rr c} ^2 ( \phi_{\rr c}^2 - \phi^2 )  }   }  }\  ,
\end{equation}
where $\phi$ is related to the adiabatic field through the relation 
\begin{equation}
\sigma(\phi) = \int_{\phi_{\rr c}} ^{\phi} \dd \phi'  \sqrt{1+ \dfrac{M^2 \phi'^2}{\phi_{\rr c} ^2 ( \phi_{\rr c}^2 - \phi'^2 )  }   }  \  .
\end{equation}
Like for an effective 1-field model, the slow-roll regime can be assumed and the slow-roll parameters can be introduced.

Let remark again that this approach is valid under the assumption that trajectories follow the above defined ellipse.  In practice, at the critical instability point, the gradient of the potential is along the $\phi$ direction. The field evolution first follow this direction and thus does not follow exactly the ellipse of minima.   Therefore the predictions are expected to be modified more or less importantly.  These modifications are studied by solving numerically the exact classical dynamics.  



\begin{figure}[h!]
 \includegraphics[width=8.0cm]{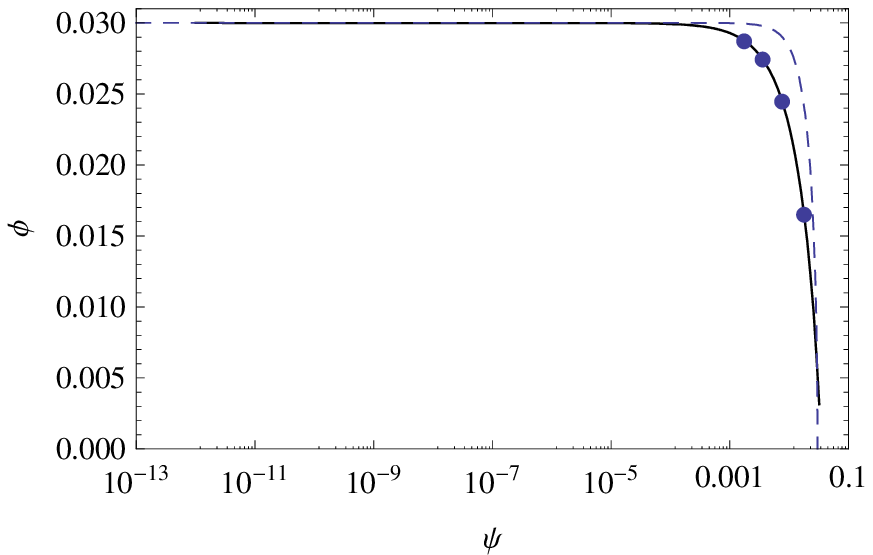}
  \includegraphics[width=8.5cm]{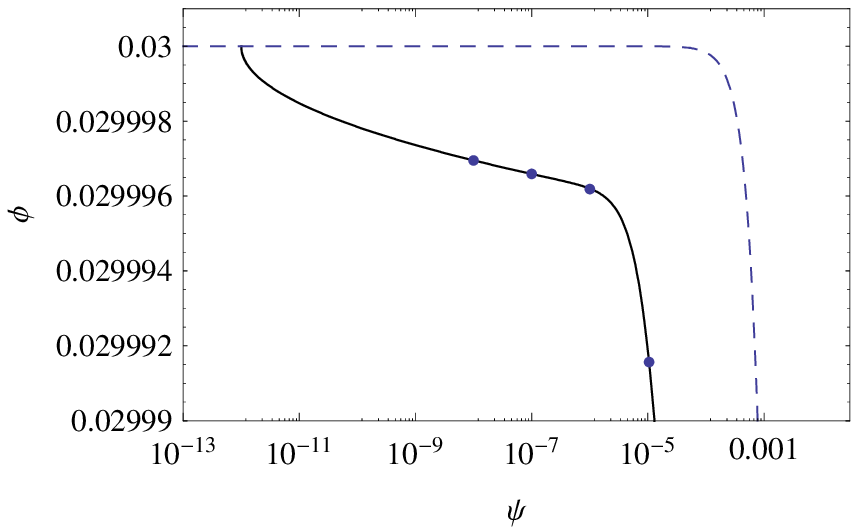}
  \includegraphics[width=8cm]{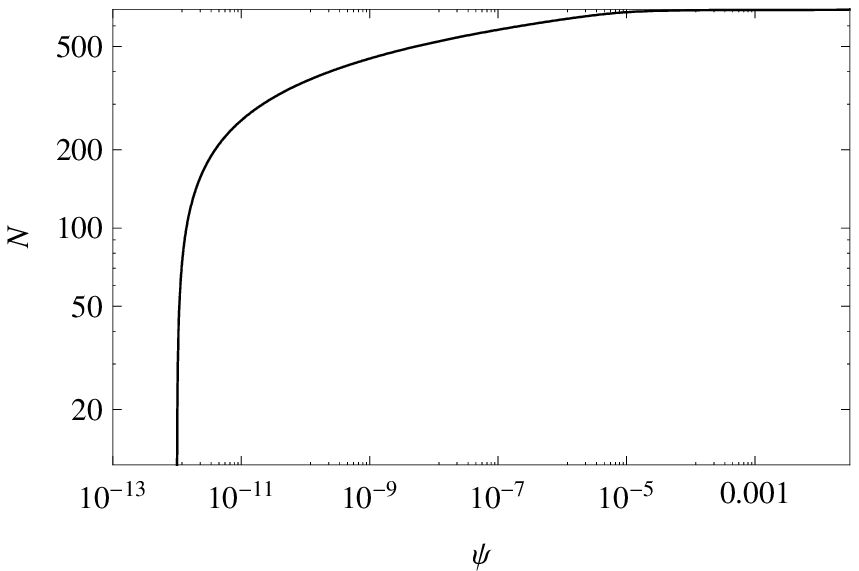}
  \caption{Top: typical trajectory (solid line) in the $(\phi , \psi) $ plane ($\phi$ and $\psi$ are in $\mpl$ units), for initial field values $\phi_i = \phi_c = 0.03 \mpl, \psi_i  = 10^{-12} \mpl $ and potential parameters $M=0.03 \mpl, \mu = 636.4 \mpl $.   The ellipse of minima of Eq.(\ref{eq:ellipse}) is also represented (dashed line).  From left to right, the points indicate where $\epsone = 10^{-3} / 10^{-2}  / 0.1  / 1 $. Center:  zoom around $\phi_{\rr c}$ for the same trajectory. From left to right, the points indicate where $n_{\rr s} = 1. / 0.97 / 0.91 / 0.65$.  Bottom:  Number of e-folds realised along this trajectory, from the critical instability point.  }
  \label{fig:traj_phipsi2}
\end{figure}

Fig. \ref{fig:traj_phipsi2}  shows a typical trajectory,  for typical potential parameters.  Starting integration at instability with $\psi_{\rr  i} = 10^{-12} \mpl$ ensures that the dynamics of the adiabatic field is not dominated by its quantum stochastic effects.  Indeed, the initial transverse displacement corresponds to the scale of quantum transverse fluctuations given by Eq.~(\ref{eq:qufluct}).  Thus for our set of parameters, one can read
\begin{equation}
\epsilon_{\rr 1} \sim \frac{\phi_{\rr c}^2 \mpl ^2}{\mu^4} > H^2 \sim 10^{-24} , 
\end{equation}
and thus quantum jumps of the adiabatic field are much smaller than classical evolution during one Hubble time.  

More than $600$ e-folds are found to be realised before inflation ends.   Therefore, 60 e-folds before the effective end of inflation, when observable modes cross the Hubble radius, instability has already been crossed and the trajectory follow a non-trivial evolution. 

The effective potential of Eq.~(\ref{eq:potadiabatic}) is thus an ideal case.  Two regimes during which inflation is possible are identified numerically: 

\begin{enumerate}
\item PHASE I:  At $\phi = \phi_c$, it is assumed that each trajectory is slightly displaced in the $\psi$ direction.  Trajectories first exhibit a non trivial behavior.  Driven by their velocity along $\phi$, they first follow the slope in the $\phi$ direction before turning, following roughly the gradient of the potential until the ellipse defined in Eq.~(\ref{eq:ellipse}) is reached.   As shown in Fig.\ref{fig:traj_phipsi2} a large number of e-folds can be  realized during this first phase.

\item PHASE II:  Trajectories reach and mostly follow the above defined ellipse.  A large number of e-folds is realized if the effective potential along the ellipse is sufficiently flat.
\end{enumerate}

Finally, if in the standard approach the tachyonic preheating begins at phase transition, in the case of waterfall inflation it can only occurs at the end of inflation, similarly to what happens for the new inflation model \cite{Desroche:2005yt}.  

\begin{figure}[h!]
\includegraphics[width=8.cm]{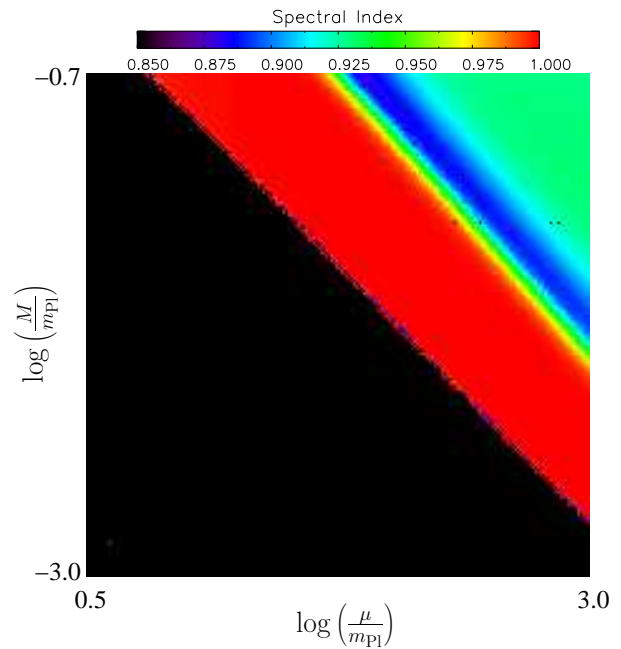}
\caption{$200 \times 200$ grid of spectral index values of the power spectrum of adiabatic perturbations, in the plane ($\mu, M$) for the exact classical dynamics, with $\phi_{\rr c} = 10^{-3} \mpl $. Black region correspond to trajectories leading to no more than $60$ e-folds after instability. }
\label{fig:nsexact}
\end{figure}

The full calculation of the primordial power spectrum, including entropic perturbations, is beyong the scope of this paper.  However, along the ellipse,  the scalar spectral index for adiabatic modes can be evaluated in the slow-roll approximation.   One has to evaluate the field value at Hubble exit of observable modes, that is when
\begin{equation}
N(\phi) = \int^{\phi} _{\phi_{\epsilon_{\rr 1} = 1 } } \dd \phi' \frac{V}{\Mpl^2 V_\sigma}  \sqrt{1+ \frac{M^2 \phi'^2}{\phi_{\rr c} ^2 ( \phi_{\rr c}^2 - \phi'^2 )  }   } \simeq 60 \  .
\end{equation}
Then the spectral index is directly determined with Eq.~(\ref{eq:slowrollparams}).   It is generically red.  
The spectral index of the adiabatic power spectrum can also be determined for numerical trajectories.  
It has been plotted along the trajectory of Fig.~\ref{fig:traj_phipsi2} as well as for a grid in the parameter space ($\mu,M$), see Fig.~\ref{fig:nsexact}.    

In this section, some waterfall trajectories leading to more than 60-folds have been found to exist.  But before to draw conclusions, it is essential to measure how generic such trajectories are in the parameter space.  This is the point of the following section, in which the full potential parameter space will be explored using a statistical MCMC method.

\section{Exploration of parameter space} \label{sec:paramspace}

The number of e-folds generated after crossing the instability point depends on the form of the potential through its three parameters $M, \mu, \nu$.  In order to draw more easily the physical interpretations, we have replaced the parameter $\nu$ by the position of the instability point $\phi_c$.   The dynamics depends also on the initial value of the auxiliary field.  It is given by the scale of its quantum fluctuations  given by Eq.~(\ref{eq:qufluct})   at instability point when it acquires a large mass.  So it is related to the potential parameter $\Lambda$ through the Friedmann-Lemaitre equation.    

To explore this 4D space, we have used a Monte-Carlo-Markov-Chains method.  Flat priors have been chosen on the logarithm of these parameters, in order to not favour any precise scale.     The chosen ranges of parameters are the following:
\begin{eqnarray}
0.3 \mpl & < & \mu < 10^4 \  \mpl \\
10^{-6}\mpl & < & M < \Mpl \\
10^{-6}\mpl & < & \phi_{\rr c} <  \Mpl \\
 10^{-60} \mpl^4 & < & \Lambda^4 < 10^{-12} \mpl^4    
\end{eqnarray}
The lower bound on $\mu$ comes from its posterior probability distribution \cite{Clesse:2009ur} to generate sufficiently long inflation inside the valley from arbitrary subplankian initial conditions.  Upper bounds on $M$ and $\phi_c$ stand because we only consider the dynamics at field values smaller than the reduced Planck mass \footnote{this restriction appears to be natural in SUSY inspired models, for which the potential is lift up at super-Planckian field values due to radiative corrections.}.   Lower bounds on $M$ and $\phi_c$ and upper bound on $\mu$ are arbitrary for numerical convenience.  Prior bounds on the parameter $\Lambda $ are such that the present constraints on the energy scale of inflation, given by nucleosynthesis and observations of the CMB, are respected.

Trajectories are integrated from the instability point.  Initial inflaton value is thus $\phi _{\rr i}= \phi_{\rr c}$.  Initial auxiliary field values are assumed to follow a gaussian distribution around $\psi = 0$, with a dispersion given by Eq.~(\ref{eq:qufluct}).
  In order to avoid strong quantum backreactions of the adiabatic field, a hard prior coming from Eq.~\ref{eq:hardprior} is enforced,
\begin{equation}
\epsilon_{\rr 1} (\phi = \phi_{\rr c}, \psi \simeq 0 ) \simeq  \frac{\phi_{\rr c}^2 \mpl^2}{\mu^4}  > \frac{H^2}{\pi \mpl^2} \sim \Lambda^4,
\end{equation}
 such that each trajectory that do not verify this condition is excluded of the Markov chain.
Trajectories are integrated until $\epsilon_{\rr 1} = 1$, the end of inflation.  The acceptance condition to a Markov chain is a realization of at least $60$ e-folds after $\phi_{\rr c}$.

\begin{figure}[h!]
\includegraphics[width=60mm]{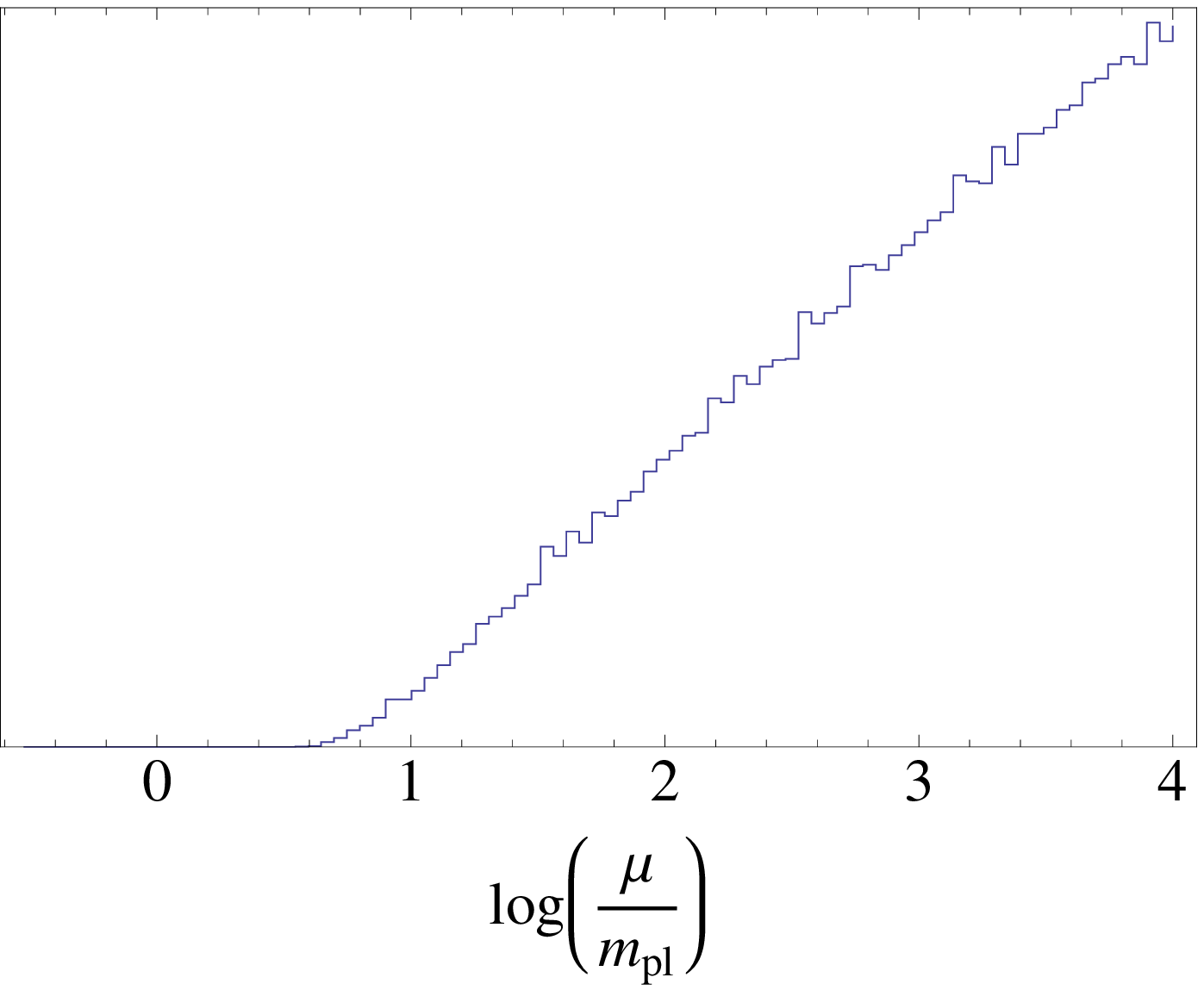}  
\raisebox{3.5mm}{
\includegraphics[width=60mm]{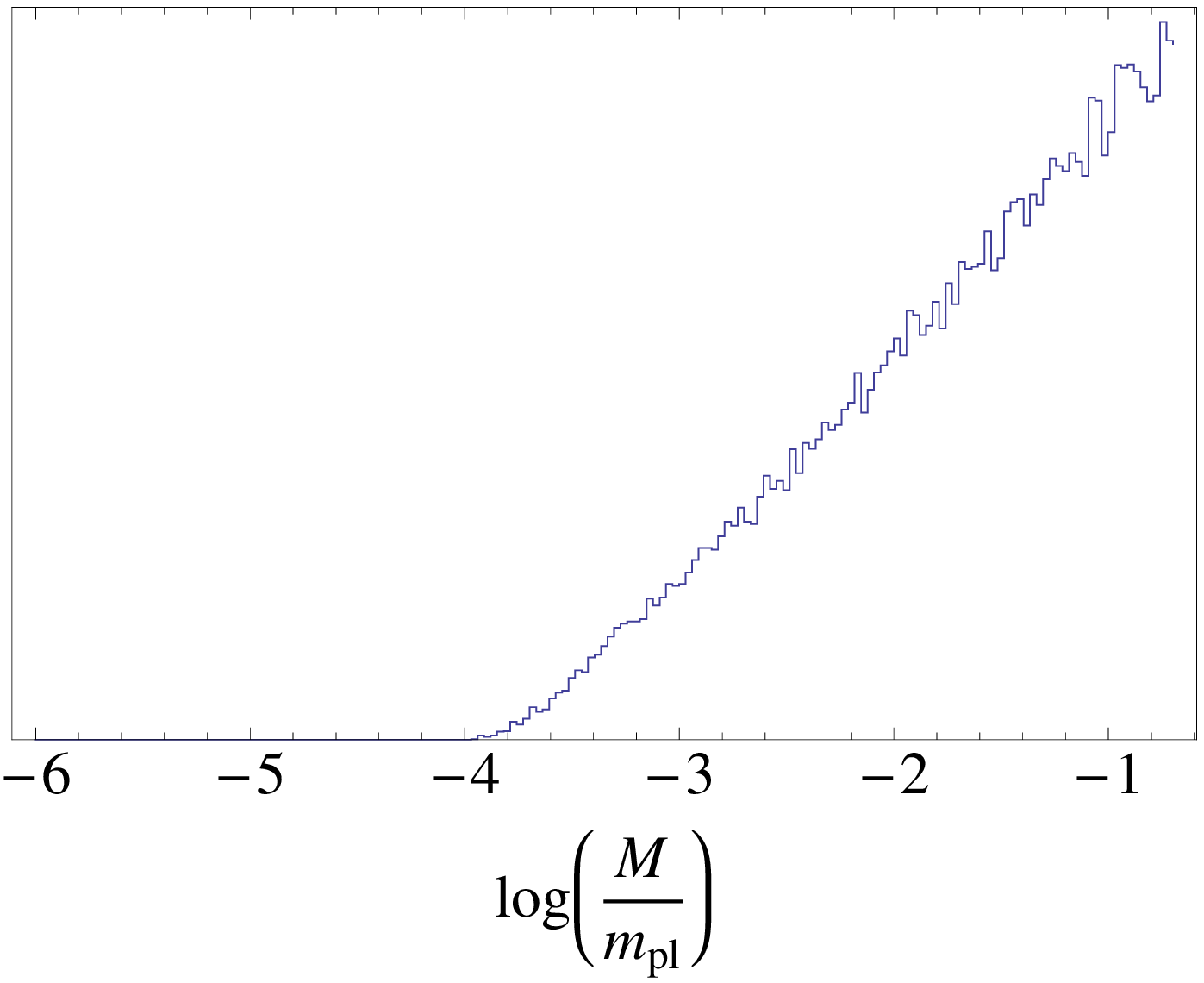} }
\caption{Marginalized posterior probability distributions of the potential parameters $\mu$ (top) and $M$ (bottom). The vertical axis is normalized such that the total area under the distribution is $1$.}
\label{fig:probamuM}
\end{figure}

\begin{figure}[h!]
\includegraphics[width=60mm]{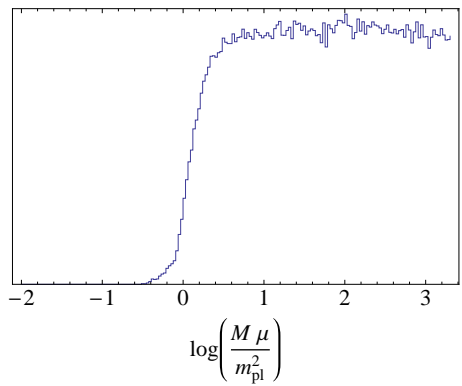}  
\caption{Marginalized posterior probability distribution of the product  $M \mu$.}
\label{fig:probaMmu}
\end{figure}

\begin{figure}[h!]
\includegraphics[width=60mm]{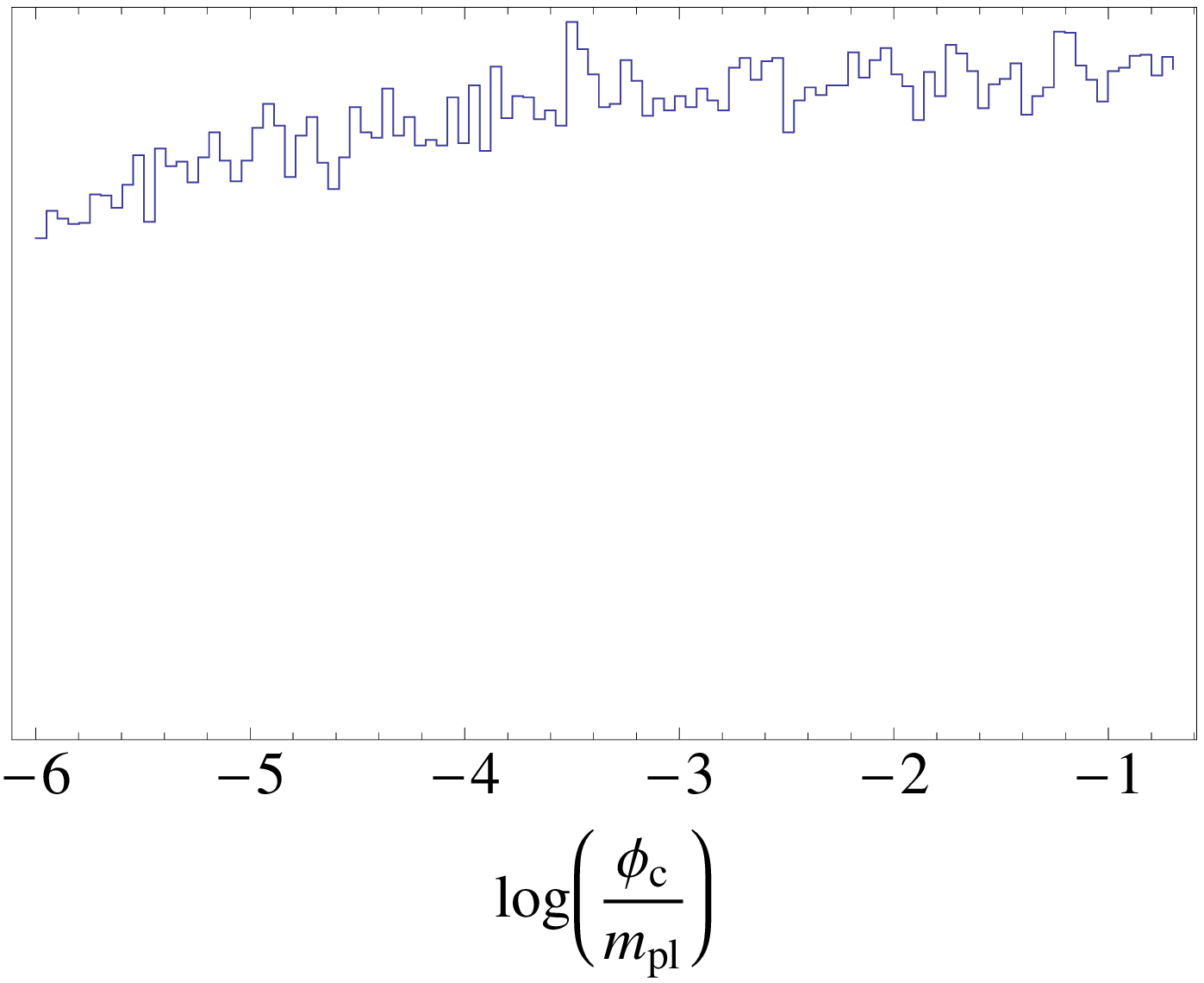}  
\raisebox{3.5mm}{
\includegraphics[width=60mm]{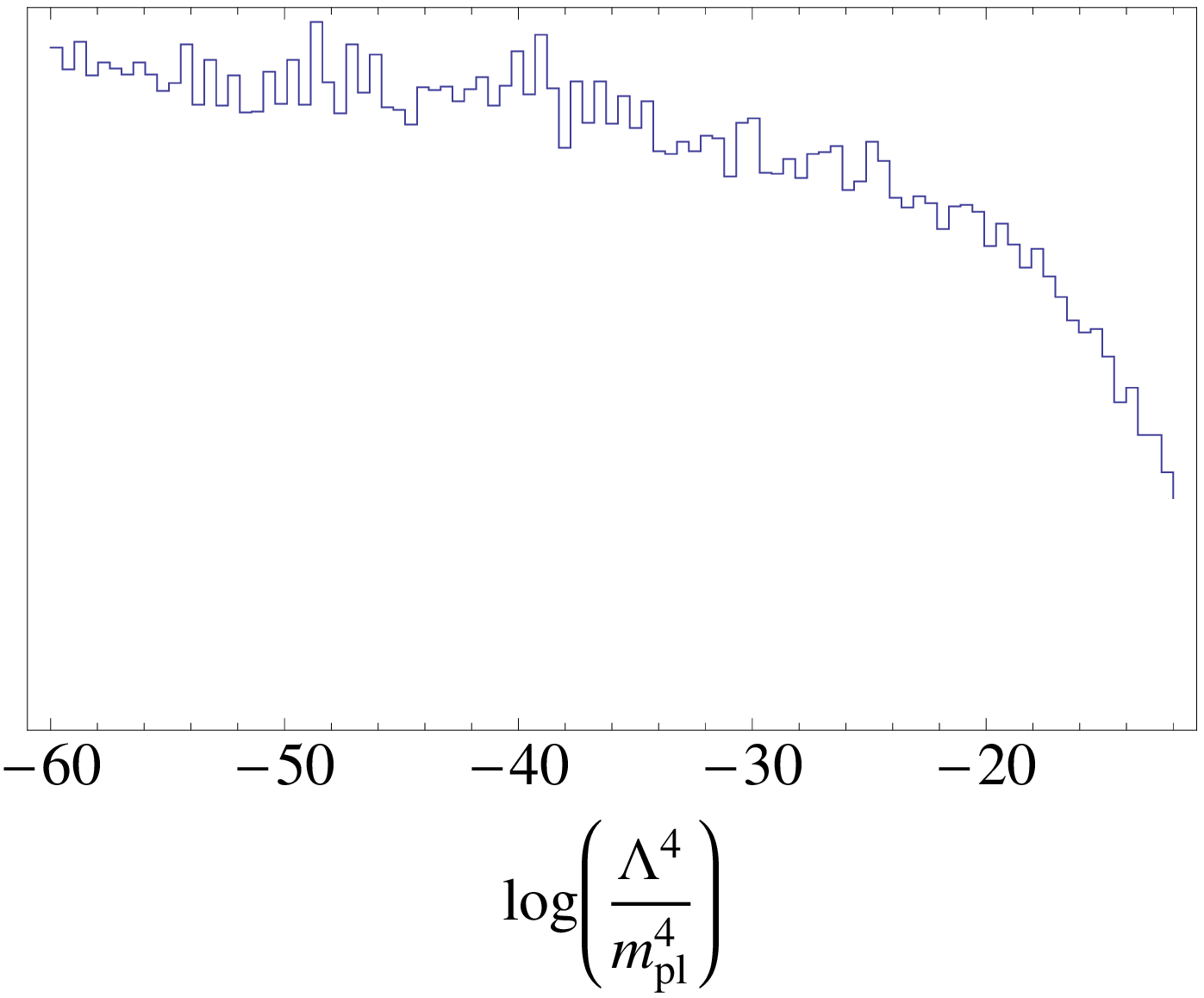} }
\caption{Marginalized posterior probability distributions of the critical point of instability  $\phi_c$ (top) and the normalizing parameter $\Lambda^4$ (bottom).}
\label{fig:probaphiclambda4}
\end{figure}


Marginalized posterior  probability density distributions, normalized such that the area under each distribution is $1$,  are shown in Figs. \ref{fig:probamuM} and \ref{fig:probaphiclambda4}.     Posterior distributions of the parameters $M$ and $\mu$ are degenerated, but the posterior distribution of their product $\mu M$  \footnote{in that case, the parameter $\mu$ is replaced by the product $M \mu$ in the MCMC simulation, using a flat prior on the log of $\mu M$.} (see Fig.\ref{fig:probaMmu}) is prior independent.   A bound on this combination is:
\begin{equation} \label{eq:boundmuM}
\log \left( \frac{ \mu M}{\mpl^2}  \right) > 0.21 \ \ 95 \% \rr{C.L.} .
\end{equation}
This bound can be rewritten $M/ \mpl \gtrsim  \mpl / \mu   $ and explained intuitively. A large number of e-folds have to be generated after the instability point, before the magnitude of the effective negative mass of the auxiliary field 
\begin{equation}
m_{\psi} (\phi) = - \sqrt 2 \frac{\Lambda^2}{M} \sqrt{ 1 - \frac{\phi^2}{\phi_{\rr c} ^2} }, 
\end{equation}
increases and becomes larger than $H$, forcing inflation to stop.  Following \cite{GarciaBellido:1996qt}, this happens in the range 
\begin{equation} \label{eq:rangephi}
\phi_{\rr c} > \phi > \phi_{\rr c} \sqrt{1- \frac{M^2}{\mpl^2}}.  
\end{equation}
During this period, the slow-roll approximation for $\phi$ is valid and the number of e-folds generated is given by Eq.~(\ref{eq:Ndephi}).  In the limit $\phi \ll \mu$,  using the range of $\phi$ obtained in Eq.~(\ref{eq:rangephi}), straightforward manipulations give $ \Delta N \sim  \mu^2 M^2 / 4 \Mpl^4$ and thus the number of e-folds is roughly fixed by the combination of the parameters $\mu$ and $M$.  From this reasoning comes also the argument that, at the end of the Phase I, $x \simeq \exp[-  M^2 /\Mpl^2 ] \sim 1 $ .

Since the analyis is restricted to the sub-planckian field dynamics, and thus to sub-planckian values of $M$, there exist a gap between
\begin{equation}
0.3 \mpl \lesssim \mu \lesssim 6 \mpl ,
\end{equation}
for which the number of e-folds generated is less than $60$.  In this regime, the slope of the potential in the $\phi$ direction is too large to generate a sufficient number of e-folds in Phase I, before the negative mass of the auxiliary field pushes the trajectory away from the $\psi = 0 $ line.

The posterior probability distribution of $\phi_{\rr c}$ is nearly flat and this parameter does not influence significatively the duration of the waterfall inflationary phase.   

The marginalized posterior probability distribution of $\Lambda^4$ is almost flat and decreases for values corresponding to the highest energy scales of inflation, without becoming negligible.  This suppression is not only due to the hard prior. 
 More important is the effect of larger initial values of the auxiliary field on the dynamics.  When classical trajectories are initially far away from the $\psi=0$ line, the phase I is reduced and less efficient to generate a large number of e-folds.

The MCMC analysis therefore provides an explicit answer to the question of how generic are trajectories realizing a large number of e-folds after the critical instability point.  They are found to occupy a large part of the parameter space, gathered in the region where Eq.~(\ref{eq:boundmuM})  is verified.   Since $\Lambda^4$ is directly linked to the energy scale of inflation, waterfall inflation is also found to be more favorable at low energy compared to the Planck scale.

Finally, let remark that the posterior probability distributions obtained with the MCMC method should be combined to the probability distributions of $\phi_{\rr i}, \psi_{\rr i}, \dot \phi_{\rr i}, \dot \psi_{\rr i}, M, \mu, \phi_{\rr c}$, obtained in~\cite{Clesse:2009ur}, related to the probability for trajectories initially outside the valley to reach the slow-roll attractor inside the inflationary valley. 

All these results stand for the original hybrid model, but the general features are expected to be reproducible with more or less efficiency, for all models in which inflation mostly occur in a nearly flat valley and end due to a tachyonic instability, like in many SUSY realizations (e.g. F-term hybrid model \cite{Dvali:1994ms}).

Notice that when Eq.~\ref{eq:boundmuM} is not verified, then the standart mechanism do work:  namely inflation stops soon after $\phi_{\rr c}$.

\section{Conclusion and discussion}

In hybrid models, the usual description assumes slow-roll inflation along a nearly flat valley ending quasi instantaneously due to a tachyonic instability triggering the tachyonic preheating \cite{Kofman:1997yn,Garcia-Bellido:1997wm, Felder:2000hj, Felder:2001kt, Copeland:2002ku, Senoguz:2004vu, Micha:2004bv,Allahverdi:2007zz}.  
  
In this paper, the waterfall phase has been studied in a regime during which inflation continues for a long time.   It has been shown that much more than $60$ e-folds can be realised classically during the waterfall, after crossing the instability point.   
Particular attention has been given to study regions in the parameter space where the classical dynamics is valid and not spoiled by quantum backreactions of adiabatic and entropic fields. 


Observable modes leave the Hubble radius when the effective potential is very flat with a negative curvature.  Instead of blue, the power spectrum of adiabatic perturbations is thus generically red.   
However, the calculation does not include the contribution of potentially observable iso-curvature modes.  This contribution was shown to be negligible in recent developments \cite{Gong:2010zf, Abolhasani:2010kn, Fonseca:2010nk, Abolhasani:2010kr, Lyth:2010ch}, but in these references a fast waterfall phase is assumed and their conclusion does not apply to the regime of potential parameters studied here.  The full numerical calculation of the primordial power spectrum should be realised soon in \cite{Clesse:2010prepa}.  The effect on the non-gaussianities produced during the tachyonic preheating \cite{Barnaby:2006cq, Barnaby:2006km} could be also important.

Therefore,  it is premature to conclude that original hybrid model is strongly disfavored by CMB experiments.   All the more since a bayesian MCMC analysis demonstrates that such trajectories are generic in a large part of the potential parameters space.  

Our result is expected to be present  in all hybrid models with an inflationary valley and a tachyonic instability.  In particular, a preliminary study of F-term supersymmetric hybrid model \cite{Dvali:1994ms} indicates that it exhibits a similar dynamics in some part of its parameter space.   

These observations may have also an important impact on questions related to the end of inflation.  In particular, if a large number of e-folds occur after symmetry breaking, the eventually formed topological defects will be diluted by expansion and thus will not affect our observable universe.   Therefore, some works constraining the schemes of symmetry breaking in grand unified theories with topological defects~\cite{Rocher:2004et} may be reviewed.   Our results should have also some impact on tachyonic preheating.  In our knowledge, all the previous studies of tachyonic preheating in hybrid inflation~\cite{Felder:2000hj, Felder:2001kt, Copeland:2002ku} neglect expansion.  Lattice simulations indicates that cosmic strings and domain walls strongly affect the way preheating phase occur.  However if such defects are diluted by a phase of inflation after symmetry breaking, lattice simulations should be updated to include expansion or to start after waterfall inflation, like in new inflation models \cite{Desroche:2005yt}.

Finally, let comment about stochastic effects.  For 1-field effective potential of hybrid inflation, these were found to not affect the classical trajectories along the valley~\cite{Martin:2005ir, Martin:2005hb}.  Authors also notice that in small field inflation, a stable solution of eternal inflation should exist at the top of the potential.   At $\phi \simeq 0$, the hybrid potential along $\psi$ reduces to a small field type and thus this observation should apply to the waterfall phase in hybrid inflation.   In particular, it would be interesting  to determine, in a full 2-field approach of stochastic effects, how the field dynamics is affected if initial conditions for the waterfall are taken along the border of the stochastic patch.

\begin{acknowledgments}
  It is a pleasure to thank J. Garcia-Bellido, D. Lyth, C. Ringeval, J. Rocher and M. Tytgat for fruitful discussions and comments. S.C. is supported by the
  Belgian Fund for research (F.R.I.A.) and the Belgian Science Policy (IAP VI-11).  
\end{acknowledgments}

\bibliography{biblio}

\begin{thebibliography}{36}
\expandafter\ifx\csname natexlab\endcsname\relax\def\natexlab#1{#1}\fi
\expandafter\ifx\csname bibnamefont\endcsname\relax
  \def\bibnamefont#1{#1}\fi
\expandafter\ifx\csname bibfnamefont\endcsname\relax
  \def\bibfnamefont#1{#1}\fi
\expandafter\ifx\csname citenamefont\endcsname\relax
  \def\citenamefont#1{#1}\fi
\expandafter\ifx\csname url\endcsname\relax
  \def\url#1{\texttt{#1}}\fi
\expandafter\ifx\csname urlprefix\endcsname\relax\def\urlprefix{URL }\fi
\providecommand{\bibinfo}[2]{#2}
\providecommand{\eprint}[2][]{\url{#2}}

\bibitem[{\citenamefont{Mazumdar and Rocher}(2010)}]{Mazumdar:2010sa}
\bibinfo{author}{\bibfnamefont{A.}~\bibnamefont{Mazumdar}} \bibnamefont{and}
  \bibinfo{author}{\bibfnamefont{J.}~\bibnamefont{Rocher}}
  (\bibinfo{year}{2010}), \eprint{1001.0993}.

\bibitem[{\citenamefont{Martin and Ringeval}(2010)}]{Martin:2010kz}
\bibinfo{author}{\bibfnamefont{J.}~\bibnamefont{Martin}} \bibnamefont{and}
  \bibinfo{author}{\bibfnamefont{C.}~\bibnamefont{Ringeval}}
  (\bibinfo{year}{2010}), \eprint{1004.5525}.

\bibitem[{\citenamefont{Clesse et~al.}(2009)\citenamefont{Clesse, Ringeval, and
  Rocher}}]{Clesse:2009ur}
\bibinfo{author}{\bibfnamefont{S.}~\bibnamefont{Clesse}},
  \bibinfo{author}{\bibfnamefont{C.}~\bibnamefont{Ringeval}}, \bibnamefont{and}
  \bibinfo{author}{\bibfnamefont{J.}~\bibnamefont{Rocher}}
  (\bibinfo{year}{2009}), \eprint{0909.0402}.

\bibitem[{\citenamefont{Clesse and Rocher}(2009)}]{Clesse:2008pf}
\bibinfo{author}{\bibfnamefont{S.}~\bibnamefont{Clesse}} \bibnamefont{and}
  \bibinfo{author}{\bibfnamefont{J.}~\bibnamefont{Rocher}},
  \bibinfo{journal}{Phys. Rev.} \textbf{\bibinfo{volume}{D79}},
  \bibinfo{pages}{103507} (\bibinfo{year}{2009}), \eprint{0809.4355}.

\bibitem[{\citenamefont{Clesse}(2009)}]{Clesse:2009zd}
\bibinfo{author}{\bibfnamefont{S.}~\bibnamefont{Clesse}}
  (\bibinfo{year}{2009}), \eprint{0910.3819}.

\bibitem[{\citenamefont{Tetradis}(1998)}]{Tetradis:1997kp}
\bibinfo{author}{\bibfnamefont{N.}~\bibnamefont{Tetradis}},
  \bibinfo{journal}{Phys. Rev.} \textbf{\bibinfo{volume}{D57}},
  \bibinfo{pages}{5997} (\bibinfo{year}{1998}), \eprint{astro-ph/9707214}.

\bibitem[{\citenamefont{Mendes and Liddle}(2000)}]{Mendes:2000sq}
\bibinfo{author}{\bibfnamefont{L.~E.} \bibnamefont{Mendes}} \bibnamefont{and}
  \bibinfo{author}{\bibfnamefont{A.~R.} \bibnamefont{Liddle}},
  \bibinfo{journal}{Phys. Rev.} \textbf{\bibinfo{volume}{D62}},
  \bibinfo{pages}{103511} (\bibinfo{year}{2000}), \eprint{astro-ph/0006020}.

\bibitem[{\citenamefont{Martin and Ringeval}(2006)}]{Martin:2006rs}
\bibinfo{author}{\bibfnamefont{J.}~\bibnamefont{Martin}} \bibnamefont{and}
  \bibinfo{author}{\bibfnamefont{C.}~\bibnamefont{Ringeval}},
  \bibinfo{journal}{JCAP} \textbf{\bibinfo{volume}{0608}}, \bibinfo{pages}{009}
  (\bibinfo{year}{2006}), \eprint{astro-ph/0605367}.

\bibitem[{\citenamefont{Linde}(1994)}]{Linde:1993cn}
\bibinfo{author}{\bibfnamefont{A.~D.} \bibnamefont{Linde}},
  \bibinfo{journal}{Phys. Rev.} \textbf{\bibinfo{volume}{D49}},
  \bibinfo{pages}{748} (\bibinfo{year}{1994}), \eprint{astro-ph/9307002}.

\bibitem[{\citenamefont{Copeland et~al.}(1994)\citenamefont{Copeland, Liddle,
  Lyth, Stewart, and Wands}}]{Copeland:1994vg}
\bibinfo{author}{\bibfnamefont{E.~J.} \bibnamefont{Copeland}},
  \bibinfo{author}{\bibfnamefont{A.~R.} \bibnamefont{Liddle}},
  \bibinfo{author}{\bibfnamefont{D.~H.} \bibnamefont{Lyth}},
  \bibinfo{author}{\bibfnamefont{E.~D.} \bibnamefont{Stewart}},
  \bibnamefont{and} \bibinfo{author}{\bibfnamefont{D.}~\bibnamefont{Wands}},
  \bibinfo{journal}{Phys. Rev.} \textbf{\bibinfo{volume}{D49}},
  \bibinfo{pages}{6410} (\bibinfo{year}{1994}), \eprint{astro-ph/9401011}.

\bibitem[{\citenamefont{Kofman et~al.}(1997)\citenamefont{Kofman, Linde, and
  Starobinsky}}]{Kofman:1997yn}
\bibinfo{author}{\bibfnamefont{L.}~\bibnamefont{Kofman}},
  \bibinfo{author}{\bibfnamefont{A.~D.} \bibnamefont{Linde}}, \bibnamefont{and}
  \bibinfo{author}{\bibfnamefont{A.~A.} \bibnamefont{Starobinsky}},
  \bibinfo{journal}{Phys. Rev.} \textbf{\bibinfo{volume}{D56}},
  \bibinfo{pages}{3258} (\bibinfo{year}{1997}), \eprint{hep-ph/9704452}.

\bibitem[{\citenamefont{Garcia-Bellido and
  Linde}(1998)}]{Garcia-Bellido:1997wm}
\bibinfo{author}{\bibfnamefont{J.}~\bibnamefont{Garcia-Bellido}}
  \bibnamefont{and} \bibinfo{author}{\bibfnamefont{A.~D.} \bibnamefont{Linde}},
  \bibinfo{journal}{Phys. Rev.} \textbf{\bibinfo{volume}{D57}},
  \bibinfo{pages}{6075} (\bibinfo{year}{1998}), \eprint{hep-ph/9711360}.

\bibitem[{\citenamefont{Felder et~al.}(2001{\natexlab{a}})}]{Felder:2000hj}
\bibinfo{author}{\bibfnamefont{G.~N.} \bibnamefont{Felder}}
  \bibnamefont{et~al.}, \bibinfo{journal}{Phys. Rev. Lett.}
  \textbf{\bibinfo{volume}{87}}, \bibinfo{pages}{011601}
  (\bibinfo{year}{2001}{\natexlab{a}}), \eprint{hep-ph/0012142}.

\bibitem[{\citenamefont{Felder et~al.}(2001{\natexlab{b}})\citenamefont{Felder,
  Kofman, and Linde}}]{Felder:2001kt}
\bibinfo{author}{\bibfnamefont{G.~N.} \bibnamefont{Felder}},
  \bibinfo{author}{\bibfnamefont{L.}~\bibnamefont{Kofman}}, \bibnamefont{and}
  \bibinfo{author}{\bibfnamefont{A.~D.} \bibnamefont{Linde}},
  \bibinfo{journal}{Phys. Rev.} \textbf{\bibinfo{volume}{D64}},
  \bibinfo{pages}{123517} (\bibinfo{year}{2001}{\natexlab{b}}),
  \eprint{hep-th/0106179}.

\bibitem[{\citenamefont{Copeland et~al.}(2002)\citenamefont{Copeland, Pascoli,
  and Rajantie}}]{Copeland:2002ku}
\bibinfo{author}{\bibfnamefont{E.~J.} \bibnamefont{Copeland}},
  \bibinfo{author}{\bibfnamefont{S.}~\bibnamefont{Pascoli}}, \bibnamefont{and}
  \bibinfo{author}{\bibfnamefont{A.}~\bibnamefont{Rajantie}},
  \bibinfo{journal}{Phys. Rev.} \textbf{\bibinfo{volume}{D65}},
  \bibinfo{pages}{103517} (\bibinfo{year}{2002}), \eprint{hep-ph/0202031}.

\bibitem[{\citenamefont{Senoguz and Shafi}(2005)}]{Senoguz:2004vu}
\bibinfo{author}{\bibfnamefont{V.~N.} \bibnamefont{Senoguz}} \bibnamefont{and}
  \bibinfo{author}{\bibfnamefont{Q.}~\bibnamefont{Shafi}},
  \bibinfo{journal}{Phys. Rev.} \textbf{\bibinfo{volume}{D71}},
  \bibinfo{pages}{043514} (\bibinfo{year}{2005}), \eprint{hep-ph/0412102}.

\bibitem[{\citenamefont{Micha and Tkachev}(2004)}]{Micha:2004bv}
\bibinfo{author}{\bibfnamefont{R.}~\bibnamefont{Micha}} \bibnamefont{and}
  \bibinfo{author}{\bibfnamefont{I.~I.} \bibnamefont{Tkachev}},
  \bibinfo{journal}{Phys. Rev.} \textbf{\bibinfo{volume}{D70}},
  \bibinfo{pages}{043538} (\bibinfo{year}{2004}), \eprint{hep-ph/0403101}.

\bibitem[{\citenamefont{Allahverdi and Mazumdar}(2007)}]{Allahverdi:2007zz}
\bibinfo{author}{\bibfnamefont{R.}~\bibnamefont{Allahverdi}} \bibnamefont{and}
  \bibinfo{author}{\bibfnamefont{A.}~\bibnamefont{Mazumdar}},
  \bibinfo{journal}{Phys. Rev.} \textbf{\bibinfo{volume}{D76}},
  \bibinfo{pages}{103526} (\bibinfo{year}{2007}), \eprint{hep-ph/0603244}.

\bibitem[{\citenamefont{Leach et~al.}(2002)\citenamefont{Leach, Liddle, Martin,
  and Schwarz}}]{Leach:2002ar}
\bibinfo{author}{\bibfnamefont{S.~M.} \bibnamefont{Leach}},
  \bibinfo{author}{\bibfnamefont{A.~R.} \bibnamefont{Liddle}},
  \bibinfo{author}{\bibfnamefont{J.}~\bibnamefont{Martin}}, \bibnamefont{and}
  \bibinfo{author}{\bibfnamefont{D.~J.} \bibnamefont{Schwarz}},
  \bibinfo{journal}{Phys. Rev.} \textbf{\bibinfo{volume}{D66}},
  \bibinfo{pages}{023515} (\bibinfo{year}{2002}), \eprint{astro-ph/0202094}.

\bibitem[{\citenamefont{Larson et~al.}(2010)}]{Larson:2010gs}
\bibinfo{author}{\bibfnamefont{D.}~\bibnamefont{Larson}} \bibnamefont{et~al.}
  (\bibinfo{year}{2010}), \eprint{1001.4635}.

\bibitem[{\citenamefont{Mazumdar}(2003)}]{Mazumdar:2003jv}
\bibinfo{author}{\bibfnamefont{A.}~\bibnamefont{Mazumdar}}
  (\bibinfo{year}{2003}), \eprint{hep-th/0310162}.

\bibitem[{\citenamefont{Gordon et~al.}(2001)\citenamefont{Gordon, Wands,
  Bassett, and Maartens}}]{Gordon:2000hv}
\bibinfo{author}{\bibfnamefont{C.}~\bibnamefont{Gordon}},
  \bibinfo{author}{\bibfnamefont{D.}~\bibnamefont{Wands}},
  \bibinfo{author}{\bibfnamefont{B.~A.} \bibnamefont{Bassett}},
  \bibnamefont{and} \bibinfo{author}{\bibfnamefont{R.}~\bibnamefont{Maartens}},
  \bibinfo{journal}{Phys. Rev.} \textbf{\bibinfo{volume}{D63}},
  \bibinfo{pages}{023506} (\bibinfo{year}{2001}), \eprint{astro-ph/0009131}.

\bibitem[{\citenamefont{Garcia-Bellido
  et~al.}(1996)\citenamefont{Garcia-Bellido, Linde, and
  Wands}}]{GarciaBellido:1996qt}
\bibinfo{author}{\bibfnamefont{J.}~\bibnamefont{Garcia-Bellido}},
  \bibinfo{author}{\bibfnamefont{A.~D.} \bibnamefont{Linde}}, \bibnamefont{and}
  \bibinfo{author}{\bibfnamefont{D.}~\bibnamefont{Wands}},
  \bibinfo{journal}{Phys. Rev.} \textbf{\bibinfo{volume}{D54}},
  \bibinfo{pages}{6040} (\bibinfo{year}{1996}), \eprint{astro-ph/9605094}.

\bibitem[{\citenamefont{Gong and Sasaki}(2010)}]{Gong:2010zf}
\bibinfo{author}{\bibfnamefont{J.-O.} \bibnamefont{Gong}} \bibnamefont{and}
  \bibinfo{author}{\bibfnamefont{M.}~\bibnamefont{Sasaki}}
  (\bibinfo{year}{2010}), \eprint{1010.3405}.

\bibitem[{\citenamefont{Fonseca et~al.}(2010)\citenamefont{Fonseca, Sasaki, and
  Wands}}]{Fonseca:2010nk}
\bibinfo{author}{\bibfnamefont{J.}~\bibnamefont{Fonseca}},
  \bibinfo{author}{\bibfnamefont{M.}~\bibnamefont{Sasaki}}, \bibnamefont{and}
  \bibinfo{author}{\bibfnamefont{D.}~\bibnamefont{Wands}}
  (\bibinfo{year}{2010}), \eprint{1005.4053}.

\bibitem[{\citenamefont{Abolhasani et~al.}(2010)\citenamefont{Abolhasani,
  Firouzjahi, and Namjoo}}]{Abolhasani:2010kn}
\bibinfo{author}{\bibfnamefont{A.~A.} \bibnamefont{Abolhasani}},
  \bibinfo{author}{\bibfnamefont{H.}~\bibnamefont{Firouzjahi}},
  \bibnamefont{and} \bibinfo{author}{\bibfnamefont{M.~H.} \bibnamefont{Namjoo}}
  (\bibinfo{year}{2010}), \eprint{1010.6292}.

\bibitem[{\citenamefont{Abolhasani and Firouzjahi}(2010)}]{Abolhasani:2010kr}
\bibinfo{author}{\bibfnamefont{A.~A.} \bibnamefont{Abolhasani}}
  \bibnamefont{and}
  \bibinfo{author}{\bibfnamefont{H.}~\bibnamefont{Firouzjahi}}
  (\bibinfo{year}{2010}), \eprint{1005.2934}.

\bibitem[{\citenamefont{Lyth}(2010)}]{Lyth:2010ch}
\bibinfo{author}{\bibfnamefont{D.~H.} \bibnamefont{Lyth}}
  (\bibinfo{year}{2010}), \eprint{1005.2461}.

\bibitem[{\citenamefont{Desroche et~al.}(2005)\citenamefont{Desroche, Felder,
  Kratochvil, and Linde}}]{Desroche:2005yt}
\bibinfo{author}{\bibfnamefont{M.}~\bibnamefont{Desroche}},
  \bibinfo{author}{\bibfnamefont{G.~N.} \bibnamefont{Felder}},
  \bibinfo{author}{\bibfnamefont{J.~M.} \bibnamefont{Kratochvil}},
  \bibnamefont{and} \bibinfo{author}{\bibfnamefont{A.~D.} \bibnamefont{Linde}},
  \bibinfo{journal}{Phys. Rev.} \textbf{\bibinfo{volume}{D71}},
  \bibinfo{pages}{103516} (\bibinfo{year}{2005}), \eprint{hep-th/0501080}.

\bibitem[{\citenamefont{Dvali et~al.}(1994)\citenamefont{Dvali, Shafi, and
  Schaefer}}]{Dvali:1994ms}
\bibinfo{author}{\bibfnamefont{G.~R.} \bibnamefont{Dvali}},
  \bibinfo{author}{\bibfnamefont{Q.}~\bibnamefont{Shafi}}, \bibnamefont{and}
  \bibinfo{author}{\bibfnamefont{R.~K.} \bibnamefont{Schaefer}},
  \bibinfo{journal}{Phys. Rev. Lett.} \textbf{\bibinfo{volume}{73}},
  \bibinfo{pages}{1886} (\bibinfo{year}{1994}), \eprint{hep-ph/9406319}.

\bibitem[{\citenamefont{Clesse and Ringeval}(2010)}]{Clesse:2010prepa}
\bibinfo{author}{\bibfnamefont{S.}~\bibnamefont{Clesse}} \bibnamefont{and}
  \bibinfo{author}{\bibfnamefont{C.}~\bibnamefont{Ringeval}}
  (\bibinfo{year}{2010}).

\bibitem[{\citenamefont{Barnaby and Cline}(2006)}]{Barnaby:2006cq}
\bibinfo{author}{\bibfnamefont{N.}~\bibnamefont{Barnaby}} \bibnamefont{and}
  \bibinfo{author}{\bibfnamefont{J.~M.} \bibnamefont{Cline}},
  \bibinfo{journal}{Phys. Rev.} \textbf{\bibinfo{volume}{D73}},
  \bibinfo{pages}{106012} (\bibinfo{year}{2006}), \eprint{astro-ph/0601481}.

\bibitem[{\citenamefont{Barnaby and Cline}(2007)}]{Barnaby:2006km}
\bibinfo{author}{\bibfnamefont{N.}~\bibnamefont{Barnaby}} \bibnamefont{and}
  \bibinfo{author}{\bibfnamefont{J.~M.} \bibnamefont{Cline}},
  \bibinfo{journal}{Phys. Rev.} \textbf{\bibinfo{volume}{D75}},
  \bibinfo{pages}{086004} (\bibinfo{year}{2007}), \eprint{astro-ph/0611750}.

\bibitem[{\citenamefont{Rocher and Sakellariadou}(2005)}]{Rocher:2004et}
\bibinfo{author}{\bibfnamefont{J.}~\bibnamefont{Rocher}} \bibnamefont{and}
  \bibinfo{author}{\bibfnamefont{M.}~\bibnamefont{Sakellariadou}},
  \bibinfo{journal}{JCAP} \textbf{\bibinfo{volume}{0503}}, \bibinfo{pages}{004}
  (\bibinfo{year}{2005}), \eprint{hep-ph/0406120}.

\bibitem[{\citenamefont{Martin and Musso}(2006{\natexlab{a}})}]{Martin:2005ir}
\bibinfo{author}{\bibfnamefont{J.}~\bibnamefont{Martin}} \bibnamefont{and}
  \bibinfo{author}{\bibfnamefont{M.}~\bibnamefont{Musso}},
  \bibinfo{journal}{Phys. Rev.} \textbf{\bibinfo{volume}{D73}},
  \bibinfo{pages}{043516} (\bibinfo{year}{2006}{\natexlab{a}}),
  \eprint{hep-th/0511214}.

\bibitem[{\citenamefont{Martin and Musso}(2006{\natexlab{b}})}]{Martin:2005hb}
\bibinfo{author}{\bibfnamefont{J.}~\bibnamefont{Martin}} \bibnamefont{and}
  \bibinfo{author}{\bibfnamefont{M.}~\bibnamefont{Musso}},
  \bibinfo{journal}{Phys. Rev.} \textbf{\bibinfo{volume}{D73}},
  \bibinfo{pages}{043517} (\bibinfo{year}{2006}{\natexlab{b}}),
  \eprint{hep-th/0511292}.

\end{thebibliography}

\end{document}